\documentclass[12pt,a4paper, amsmath,amssymb]{article}
\usepackage[margin=0.8in]{geometry}
\usepackage{amssymb}
\usepackage{amsmath}
\usepackage{amsfonts}
\usepackage{bm}
\usepackage[utf8]{inputenc}
\usepackage{graphicx}
\usepackage{float}
\usepackage{multirow} 
\usepackage[colorlinks,linkcolor = blue,
urlcolor  = blue,
citecolor = blue,
anchorcolor = blue]{hyperref}
\usepackage{booktabs}

\usepackage{xcolor}
\usepackage{ctable} 
\usepackage{adjustbox}
\def\thefootnote{\fnsymbol{footnote}}
\usepackage{cite}

\begin{document}
	{
\begin{center}
{\Large \textbf{Responses of AdS black holes to the collective influence of quintessence and string cloud }}
\thispagestyle{empty}
\vspace{1cm}

{\sc
	
	H. Laassiri$^1$\footnote{\url{hayat.laassiri@gmail.com}},
	A. Daassou$^1$\footnote{\url{ahmed.daassou@uca.ma}},
	R. Benbrik$^1$\footnote{\url{r.benbrik@uca.ac.ma}}\\
}
\vspace{1cm}
{\sl
	$^1$Fundamental and Applied Physics Laboratory, Physics Department, Polydisciplinary Faculty,
	Cadi Ayyad University, Sidi Bouzid, B.P. 4162, Safi, Morocco.\\
}
\end{center}
\vspace*{0.1cm}
\begin{abstract}
This research paper explores the interplay between quintessence and a cloud of strings, focusing on their influence on critical points, behaviors, and fractional-order phase transitions in AdS black holes. We analyze the thermodynamic properties of AdS black holes surrounded by quintessence and a cloud of strings, with particular attention to the effects of thermal fluctuations on their thermodynamics. Using critical conditions, we derive approximate analytical expressions for the critical points applicable to all types of these black holes.  
We further study the impact of the quintessence parameter and the intensity of the cloud of strings on critical behaviors, employing three-dimensional visualizations to depict these effects for different values of \(\omega_q\). Our results show how these factors affect the locations of phase transitions and the regions where both phases coexist. In addition, we examine fractional-order phase transitions, which reveal significant transformations in the black hole system.

\end{abstract}


\def\thefootnote{\arabic{footnote}}
\setcounter{page}{0}
\setcounter{footnote}{0}

\newpage
\section{Introduction}
\label{intro}\
Highly accurate astronomical observations concerning the evolution of the universe have led to the conclusion that there exists an energy component exerting a repulsive gravitational effect and associated with negative pressure. This energy is responsible for the accelerated expansion of the universe. This naturally raises the question of the origins of this negative pressure. One potential explanation is the presence of a ubiquitous cosmic fluid, often referred to as quintessence dark energy, believed to drive this cosmic expansion, as indicated by various Refs.~\cite{1,2}. It's worth mentioning that within an astrophysical framework, quintessence is expected to significantly influence black holes, particularly in their vicinity. 
Another possible candidate for the energy driving the universe's accelerated expansion is the cosmological constant. This constant maintains spatial homogeneity and possesses a fixed value, holding relevance in specific astrophysical contexts~\cite{3}. Taking into account the notion that the universe's fundamental nature may be better comprehended as a collection of extended entities, such as one-dimensional strings rather than infinitesimal point particles, Letelier devised a formalism~\cite{4} to explore black holes surrounded by a spherically symmetric string cloud. The presence of this string cloud significantly modifies the black hole's horizon, distinguishing it from the Schwarzschild black hole. 
In Ref.~\cite{5}, the authors concluded that the Hubble constant $H_0$ is reduced relative to $\Lambda$CDM in any quintessence model with an exponential coupling.

The critical phenomena associated with charged AdS black holes have been extensively studied, making significant contributions to our understanding~\cite{6,7,8,9,10,11,12,13,14,15,16,17,18,19,20,21,22,23,24,25,26,27,28,29,30,31}. In Ref.~\cite{32}, a detailed analysis of phase transitions in charged AdS black holes under Gauss-Bonnet gravity revealed a resemblance to the liquid-to-gas transition in van der Waals fluids, primarily due to the influence of quintessence. 
Quintessence's impact on the criticality of pressure-volume relations in RN-AdS black holes was explored in Ref.~\cite{33}, which concluded that it does not fundamentally alter the small/large black hole phase transition. Utilizing the Maxwell equal-area law, Ref.~\cite{34} demonstrated similarities between black hole phase transitions  influenced by quintessence and van der Waals fluid behavior. Similarly, the phase transition of quintessential Kerr-Newman-AdS black holes, analyzed using the Maxwell equal-area law~\cite{35,36}, exhibited parallels with vdW systems below a critical temperature. 
A recent advancement~\cite{37} presented a generalized solution to Einstein's field equations incorporating dark energy, while our recent work~\cite{38} investigated the impact of dark energy on AdS black holes, yielding a Kerr-Newman-AdS black hole model encompassing both quintessence and a spherically symmetric cloud of strings. 
These studies emphasize the intricate interplay between various components in describing black holes and their connections to broader astrophysical and cosmological phenomena, thereby advancing our understanding of the universe's mechanisms.

This work primarily focuses on examining the collective impact of dark energy and a cloud of strings on critical phenomena within AdS black holes. The paper's structure can be outlined as follows: In Section \ref{:2}, we provide an overview of the thermodynamic properties of AdS black holes in the presence of these two additional sources of energy. We also investigate how correction terms impact the stability of these black holes. Transitioning to Section \ref{:3}, we derive and analyze various critical thermodynamic parameters, including temperature, pressure, volume, entropy, and Gibbs free energy. These calculations are expressed in terms of several variables, including the charge $(Q)$, angular momentum $(j)$, the quintessential state parameter $\omega_q$, a constant $(b)$ representing the presence of the cloud of strings, and $\alpha$, a positive parameter measuring the intensity of the quintessential field around the black hole. Section \ref{:4} provides an extensive analysis of critical behaviors, employing a three-dimensional perspective. This section illustrates how the combined presence of the cloud of strings and quintessence influences the locations of phase transition points and the regions where two phases coexist. Moving to Section \ref{:5}, we focus on the computation and discussion of fractional-order phase transitions, which indicate transformations within the system. Finally, the paper concludes with a summary of our findings and an overview of the implications of our results in Section \ref{:6}.
\section{The consequences of thermal fluctuations on the thermodynamics of AdS black holes in the presence of quintessence and a string cloud}
\label{:2}
In a prior study~\cite{39}, the metric for a charged rotating AdS black hole surrounded by a combination of strings and quintessence was presented. In this paper, we describe the thermodynamic properties of AdS black holes influenced by these two additional sources of energy\\
\begin{equation}{\Delta_{r} =a^2+\left(1-\frac{8}{3} a^2 P \pi \right)^2 Q^2-2 M r+(1-b) r^2+\frac{8}{3} P \pi  r^2 \left(a^2+r^2\right)-r^{1-3 \omega
		_q} \alpha },\label{1}\end{equation}
the symbols are assigned the following interpretations: \( a \) represents the angular momentum per unit mass of the black hole, \(\omega_q\) denotes the quintessential state parameter, and \( b \) is a constant introduced to account for the existence of the string cloud. We derive \( M \), the system's physical mass, by using \( \Delta_{r} = 0 \)
\begin{equation}
{M =\frac{ \left( \left(3-8 a^2 P \pi \right)^2 Q^2+3a^2 \left(3+8 P \pi  r^2\right)+3r \left(-3 (-1+b) r+8 P \pi  r^3-3 r^{-3 \omega
			_q} \alpha \right)\right)}{2 \left(3-8 a^2 P \pi \right)^2 r}}.
\label{2}
\end{equation}
Using the expression \( \dfrac{\Delta'_{r}}{4\pi(a^{2}+r^{2})} \), the Hawking temperature can be expressed as
\begin{equation}
{T=-\frac{a^2+\frac{1}{9} \left(3-8 a^2 P \pi \right)^2 Q^2+(-1+b) r^2-\frac{8}{3} P \pi  r^2 \left(a^2+3 r^2\right)-3 r^{1-3 \omega
			_q} \alpha  \omega _q}{4 \pi  r \left(a^2+r^2\right)}}.\label{3}
\end{equation}
The entropy of a black hole is directly proportional to the surface area of its event horizon and can be expressed accordingly
\begin{equation}
{S=\dfrac{3 \pi  \left(a^2+r^2\right)}{3-8 a^2 P \pi }}.\label{4}
\end{equation}
To examine how thermal fluctuations affect the thermodynamics of black holes surrounded by quintessence and a cloud of strings, we construct graphs illustrating the corrected entropy for a range of correction parameters. The expression for this adjusted entropy is detailed in Ref.~\cite{30}. Our calculations are based on the canonical partition function, which forms the fundamental basis for our computations.
\begin{align}\label{ 5}
{Z(\beta ) = \int _0^{\infty }\dfrac{\sigma (E)}{\exp ^{\text{$\beta $E}}}\text{dE}},
\end{align}
the symbol \( \sigma(E) \) represents the quantum density of the system. By applying the inverse Laplace transformation, and following an extensive and rigorous calculation process, we derive the density of states
\begin{align}
{\sigma (E) = \dfrac{-i e^S}{2\pi}\int _{n-\text{i$\infty $}}^{n+\text{i$\infty $}}\exp ^{\dfrac{(\beta -n)^2}{2}\dfrac{d^2S_0}{d^2\beta
	}}\text{d$\beta $}},\label{6}
\end{align}
using $ S_{0} $ to denote the corrected entropy, it is expressed as follows
\begin{align}
 {S_0 = S - \beta \ \text{Ln}\left(S \ T^2\right) + \dfrac{\beta _0}{S}},\label{Eq.7} 
\end{align}
utilizing correction parameters $\beta$ and $\beta_{0}$, in accordance with Eq. (\ref{3}) and Eq. (\ref{4}), we can derive the results, including the standard logarithmic corrections when $\beta_{0}$ is set to zero\ 
\begin{equation}
\begin{aligned}
S_0=&\dfrac{3 \pi  \left(a^2+r^2\right)}{3-8 a^2 P \pi }-\beta  \ \text{Ln}\left(3 \left(a^2+\frac{1}{9} \left(3-8 a^2 P \pi \right)^2
	Q^2+(-1+b) r^2-\frac{8}{3} P \pi  r^2 \left(a^2+3 r^2\right) \right.\right. \\	
& \left.\left.-3 r^{1-3 \omega _q} \alpha \  \omega _q\right){}^2/\left(16 \pi  \left(3-8 a^2 P \pi \right) r^2 \left(a^2+r^2\right)\right)\right).
 \end{aligned}
\end{equation}

\begin{figure}[H]
	\centering
	\includegraphics[width=0.6\linewidth, height=10cm]{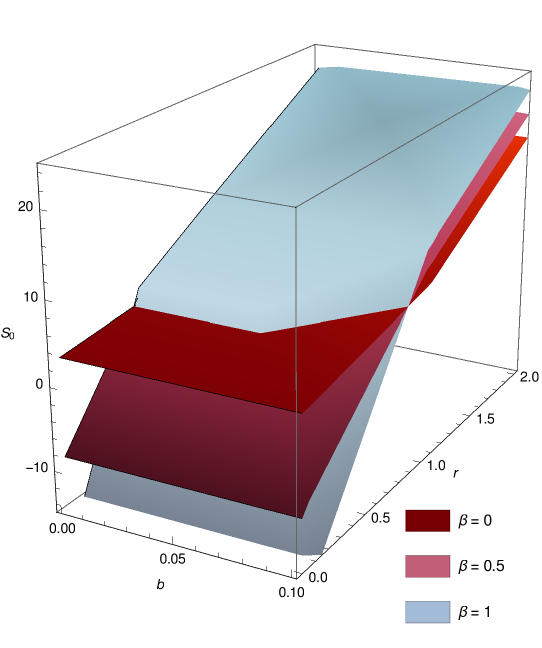}
	\caption{Dual entropy dynamics: Examining $ b $ and $ r $ influences on corrected and uncorrected entropy with $ j = Q = 1 $ and $ \alpha= 0.06 $.}
	\label{fig1}
\end{figure}
Figure \ref{fig1} depicts the changes in corrected and uncorrected entropy with respect to the parameter $r$, while maintaining constant values for $j$, $Q$, $\alpha$, and $\omega_q$ at 1, 1, 0.06, and -1/3, respectively, across different values of $b$. The graph consistently illustrates that the entropy at equilibrium ($\beta=0$) remains positive and steadily increases, adhering to the second law of black hole thermodynamics, which asserts that black hole entropy always increases.

Nonetheless, when the correction parameter $\beta$ deviates from zero, smaller black holes exhibit a tendency toward negative entropy. The influence of thermal fluctuations becomes especially pronounced with higher values of $\beta$, extending across various parameters of the string cloud $b$. Furthermore, for larger horizon radius, the corrected entropy consistently adopts positive values, mirroring the pattern observed in the uncorrected entropy. This finding carries significant implications, emphasizing the substantial impact of thermal fluctuations on the thermodynamics of smaller black holes and underscoring the intricate relationship with varying correction parameter and horizon radius.
\section{Exploring the analytical study of critical points in AdS black holes with quintessence and a string cloud}
\label{:3}

The critical points can be determined by satisfying the condition $ \partial P_v = \partial P_{v,v} = 0 $. In this context, the pressure $ P $ must be expressed in terms of variables $ v $, $ T $, $ j $, $ Q $, $ \alpha $, $ b $, and $ \omega_q $. From the expression $ j = a M $, we can derive the angular momentum per unit mass using Eq. (\ref{2})
\begin{equation}
{a=\dfrac{6 j r^{1+3 \omega _q}}{3 Q^2 r^{3 \omega _q}+3 r^{2+3 \omega _q}-3 b r^{2+3 \omega _q}+8 P \pi  r^{4+3 \omega _q}-3 r \alpha
}}.\label{Eq.9}
\end{equation}
Our analysis truncates the Taylor series expansion of the black hole angular momentum to second order. This choice was not arbitrary but based on numerical considerations. Specifically, our numerical calculations show that including higher-order terms (beyond the second order) does not yield critical values that are consistent with the expected results when all additional parameters in the present work are set to zero. Thus, truncating at second order ensures consistency with established critical values and maintains the physical reliability of the model. Finally, the second-order approximation proves to be suitable for our approach.
The perturbative approach we employ is carefully chosen to balance mathematical tractability and physical relevance. The second-order truncation is sufficient to capture the key features of the system without overcomplicating the analysis. We specifically focus on perturbations that respect the constraints of the black hole under consideration.

Using the previous expression for the angular momentum per unit mass, the physical volume can be expressed in the following manner
\begin{equation}
{V=\dfrac{\partial M}{\partial P}={\frac{\pi  v^3}{6}=}{\frac{4 \pi  r^3}{3}+\frac{16 \pi  r^{2+3 \omega _q} \left(9 r^{1+3 \omega _q}-6 b r^{1+3 \omega _q}+32 P \pi  r^{3+3 \omega _q}-6
			\alpha \right) j^2}{\left(3 Q^2 r^{3 \omega _q}+3 r^{2+3 \omega _q}-3 b r^{2+3 \omega _q}+8 P \pi  r^{4+3 \omega _q}-3 r \alpha \right){}^2}}+O[j]^3,}\label{10}
\end{equation}
interpreting $ v $ as the specific volume and making use of Eqs. (\ref{3}), (\ref{Eq.9}), and (\ref{10}), the pressure can be formulated as follows
\begin{equation}
\begin{aligned}
{P=\dfrac{v^{-3 \left(2+\omega _q\right)} \left(3 v^2 P_1+64 j^2 v^{6 \omega _q} \left(v^{3 \omega _q} P_2+3\ 2^{1+3 \omega _q} v \alpha
		\left(P_3-3\ 2^{2+3 \omega _q} v^3 \alpha  \omega _q^2\right)\right)\right)}{6 \pi  \left(-4 Q^2 v^{3 \omega _q}+(-1+b) v^{2+3 \omega _q}+2^{1+3
			\omega _q} v \alpha \right){}^3}}.\label{11}
\end{aligned}
\end{equation}
The critical condition for charged AdS black holes surrounded by quintessence and a cloud of strings, achieved by setting $ j=0 $, can be expressed as follows
\begin{equation}
{v^{-5-3 \omega _q} \left(-v^{3 \omega _q} \left(8 Q^2+v^2 (-1+b+\pi  T v)\right)+9\ 8^{\omega _q} v \alpha  \omega _q+9\ 8^{\omega
		_q} v \alpha  \omega _q^2\right)= 0,} \label{12}
\end{equation}
\begin{equation}
{v^{-3 \left(2+\omega _q\right)} \left(v^{3 \omega _q} \left(40 Q^2+v^2 (-3+3 b+2 \pi  T v)\right)-9\ 2^{2+3 \omega _q} v \alpha  \omega _q-63\
	8^{\omega _q} v \alpha  \omega _q^2-27\ 8^{\omega _q} v \alpha  \omega _q^3\right)= 0,} \label{13}
\end{equation}
the solution of Eq. (\ref{12}) and Eq. (\ref{13}) allows us to determine the analytical expression of the critical point for any value of omega. For {$ \omega _q = -2/3 $}, we have obtained
$$
{T_{c} = \frac{\sqrt{6}-2 \sqrt{6} b+\sqrt{6} b^2-9 \sqrt{1-b} Q \alpha }{18 \sqrt{1-b} \pi  Q},\hspace{1.5cm}V_{c} = \frac{2 \sqrt{6} Q}{\sqrt{\left(1-b\right)}},
	\hspace{1.5cm}P_{c}  =\frac{(1-b)^2}{96
		\pi  Q^2}}, $$
$${S_{c}=\frac{6 \pi  Q^2}{\sqrt{(-1+b)^2}}},\hspace{3cm}{G_{c}={\sqrt{\frac{2}{3}} \sqrt{\left(1-b \right)}}Q}.$$

With {$ \omega _q = -1/3 $}, the analytical expression of the critical point is expressed as follows 
$${T_{c} = \frac{\sqrt{6}\left({1-b-\alpha }\right)^{3/2}}{18 \pi  Q},\hspace{1.5cm}V_{c} = \frac{2 \sqrt{6} Q}{\sqrt{\left(1-b-\alpha \right)}}
	,\hspace{1.5cm}P_{c} = \frac{(1-b-\alpha )^2}{96 \pi
		Q^2}},$$
$${S_{c}=\frac{6 \pi  Q^2}{\sqrt{(-1+b+\alpha )^2}},\hspace{1.5cm}{G_{c} =\sqrt{\frac{2}{3}} Q \sqrt{\left(1-b-\alpha \right)}}}.$$
To obtain the analytical expression of the critical point for rotating AdS black holes under the combined effects of quintessence and a string cloud for any value of $ \omega _q $, we enforce the condition $ Q = 0 $ in Eq. (\ref{11}) and apply the critical condition. For {$ \omega _q = -1/3 $}, we derived
$${T_{c} = \frac{3\ 37^{3/4} ((-1+b+\alpha )^2)^\frac{7}{8}}{242 \sqrt{2} \sqrt{j} \pi },\hspace{1.5cm}V_{c} = \frac{2 \sqrt{2}\ \ 37^{1/4} \sqrt{j}}{\left(\sqrt{(-1+b+\alpha )^2}\right)^{3/4}},}$$
$$\hspace{1.5cm} {S_{c} =  {\frac{3 \sqrt{10} j \pi }{\left((1-b-\alpha )^2\right)^{1/4}}}},\hspace{1.5cm} {G_c=M_{1c}-S_c T_c},$$\
$$	{P_{c} = {\frac{\sqrt{\left((-1+b+\alpha )^2\right)^{3/2}} \left(-13875+25610 \sqrt{(-1+b+\alpha )^2}+3 b \ \eta +3 \alpha \ \eta \right)}{214896
			\sqrt{37} j \pi }}},$$ \
with $${\eta =\left(4625-121 \sqrt{(-1+b+\alpha )^2}\right)}.$$
The analytical critical points for Kerr-Newman AdS black holes can be effectively determined by satisfying the conditions $\frac{\partial T}{\partial S} = \frac{\partial^2 T}{\partial S^2} = 0$. In this scenario, we can express the relationship between $r$ and the variables $a$, $P$, and $S$ using Eq. (\ref{4}) as follows
\begin{equation}
{r =  \sqrt{\frac{-3 a^2 \pi +3 S-8 a^2 P \pi  S}{3 \pi }}},\label{14}
\end{equation}
afterward, we can describe the parameter $a$ using the expression $a = \frac{M}{j}$ in conjunction with Eq. (\ref{2}) and Eq. (\ref{14})
\begin{equation}{a = \dfrac{6 j \sqrt{\pi } S^{\frac{1}{2}+\frac{3 \omega _q}{2}}}{3 \pi  Q^2 S^{\frac{3 \omega _q}{2}}+3 S^{1+\frac{3 \omega _q}{2}}-3
		b S^{1+\frac{3 \omega _q}{2}}+8 P S^{2+\frac{3 \omega _q}{2}}-3 \pi ^{\frac{1}{2}+\frac{3 \omega _q}{2}} \sqrt{S} \alpha }},
\label{15}
\end{equation}
by substituting the previously mentioned Equations (Eq. (\ref{14}) and Eq. (\ref{15})) into Eq. (\ref{3}), we arrive at the following result
\begin{equation}
\begin{aligned}
{T=\frac{S^{-\frac{3}{2} \left(1+\omega _q\right)} \left(T_1 T_2+6 j^2 \pi ^2 S^{3 \omega _q} T_3\right)}{4 \sqrt{\pi } \left(3 \pi
		Q^2 S^{\frac{3 \omega _q}{2}}+S^{1+\frac{3 \omega _q}{2}} (3-3 b+8 P S)-3 \pi ^{\frac{1}{2}+\frac{3 \omega _q}{2}} \sqrt{S} \alpha \right){}^2}}.
\end{aligned}\label{16}
\end{equation} 
The analytical critical points, given {$\omega_q = -1/3$}, are expressed as follows	\
	
${S_{c} ={\dfrac{3 \pi  \left(Q^2+\beta_{1} \right)}{\sqrt{(1-b-\alpha )^2}},}}\hspace{1.5cm}
{ { P _{c} = {\dfrac{Q^4 \left(-Q^2+\beta_{1} \right)-5 j^2 (-1+b+\alpha ) \left(-3 Q^2+2 \beta_{1} \right)}{3600 j^4 \pi },}}} $

{$ T _{c}=\left(-9 \pi ^3 \left(Q^2+3 \left(-1+b+\alpha -\beta _2\right) \beta _3\right) \left(Q^2+\left(-3 (-1+b+\alpha )+\beta _2\right)
	\beta _3\right){}^2+6 j^2 \pi ^2\right. $}\

{$ \left.\left. \left(\pi  Q^2 \left(-3+\beta _2\right)-3 \pi  \left(-3 (-3+b+\alpha )+(6+b+\alpha ) \beta _2+\beta _2^2\right) \beta _3\right)\right)\right/(108 \sqrt{3} \pi ^4 \beta _3^{3/2}$}\

{$ \left. \left(Q^2+\left(-3 (-1+b+\alpha )+\beta _2\right) \beta _3\right){}^2\right) $},\

$ {G_c=M_{2c}-S_c T_{_c}.} $\\

For $\omega_q = -2/3$, the expression of the analytical critical point is given by\

${{S_c=\dfrac{3 \left(\pi ^{3/2} Q^2+2 \alpha +\sigma \right)}{\sqrt{(1-b)^2} \sqrt{\pi }}}},$\

${{P_{c} =\dfrac{\left(5 (-1+b) j^2 \pi ^3 \left(3 \pi ^{3/2} Q^2+6 \alpha -2 \sigma \right)+\left(\pi ^{3/2} Q^2+2 \alpha \right)^2 \left(-\pi
			^{3/2} Q^2-2 \alpha +\sigma \right)\right)}{3600 j^4 \pi ^{11/2}}}},$\
		 
$ {T_c=\left(6 j^2 \pi ^3 \left(12 \alpha  S_c^{3/2}+\pi ^{3/2} Q^2 \left(-3+8 P_c S_c\right)-\sqrt{\pi } S_c \left(b \left(-3+8 P_c
	S_c\right)+\left(3+8 P_c S_c\right){}^2\right)\right)\right.} $\

$ {+\left(-2 \alpha  S_c^{3/2}-\sqrt{\pi } \left(\pi  Q^2+S_c \left(-1+b-8 P_c S_c\right)\right)\right) \left(-3 \alpha  S_c^{3/2}+\sqrt{\pi }
	\left(3 \pi  Q^2+S_c (3-3 b+\right.\right.} $\

$ {\left.\left.\left.\left.8 P_c S_c\right)\right)\right){}^2\right)/\left(4 \pi  S_c^{3/2} \left(-3 \alpha  S_c^{3/2}+\sqrt{\pi } \left(3 \pi
	Q^2+S_c \left(3-3 b+8 P_c S_c\right)\right)\right){}^2\right)}, $\

$ {G_c= M_{3c}-S_c T_c}.$\

To confirm our results and approach, 
		we define the relative deviation for a thermodynamic quantity \( \Delta T_c = \frac{T_c - T_{ac}}{T_c} \), where \( T_{ac} \) is the critical point obtained from our analysis for rotating AdS black holes (with the additional parameters \( b, \alpha, \omega_q \) set to zero), and \( T_c \) is the critical temperature obtained from Ref.~\cite{40}, defined as \( T_c = \frac{0.041749}{\sqrt{j}} \). This definition enables us to quantify the accuracy of our results relative to the known analytical expression.
	
	The results of our analysis are summarized in Table~\ref{tab:critical_temperatures}, which shows the critical temperature values \( T_{ac} \) and relative deviations (\(\Delta T_c\)) for various orders of expansion (\(\mathcal{O}(j^3)\), \(\mathcal{O}(j^5)\), \(\mathcal{O}(j^7)\)) at different values of the rotation parameter \( j \).
	
	\begin{table}[h!]
		\centering
		\renewcommand{\arraystretch}{1.1}
		\setlength{\tabcolsep}{4pt}
		\begin{tabular}{|c|c|c|c|}
			\hline
			\textbf{\(j\)} & \textbf{Order} & \textbf{\(T_{ac}\)} & \textbf{\(\Delta T_c\) (\%)} \\ 
			\hline
			\multirow{3}{*}{0.01} 
			& \(\mathcal{O}(j^3)\) & 0.417049 & 0.106 \\ 
			& \(\mathcal{O}(j^5)\) & 0.416762 & 0.174 \\ 
			& \(\mathcal{O}(j^7)\) & 0.416754 & 0.176 \\
			\hline
			\multirow{3}{*}{0.1} 
			& \(\mathcal{O}(j^3)\) & 0.131882 & 0.105 \\ 
			& \(\mathcal{O}(j^5)\) & 0.131791 & 0.174 \\ 
			& \(\mathcal{O}(j^7)\) & 0.131789 & 0.176 \\ 
			\hline
			\multirow{3}{*}{1} 
			& \(\mathcal{O}(j^3)\) & 0.041704 & 0.108 \\ 
			& \(\mathcal{O}(j^5)\) & 0.041676 & 0.175 \\ 
			& \(\mathcal{O}(j^7)\) & 0.041675 & 0.177 \\
			\hline
		\end{tabular}
		\caption{Critical temperature values \( T_{ac} \) and relative deviations (\(\Delta T_c\)) for various orders of expansion.}
		\label{tab:critical_temperatures}
	\end{table}
	
	As shown in our results, the relative deviations (\(\Delta T_c\)) slightly increase with the inclusion of higher-order terms, as presented in Table~\ref{tab:critical_temperatures}. For instance, for \( j = 0.01 \), the relative deviation rises from 0.106\% at \(\mathcal{O}(j^3)\) to 0.174\% at \(\mathcal{O}(j^5)\), and further to 0.176\% at \(\mathcal{O}(j^7)\). This trend indicates that the lower-order terms provide results that are more accurate and closer to the analytical reference values compared to higher-order terms. This behavior can be attributed to the perturbative nature of the expansion, where the dominant contributions are captured at lower orders, while higher-order terms, despite including additional corrections, introduce increasingly negligible effects or compounded uncertainties.
	
	Our conclusion that higher-order effects do not reveal any novel physical phenomena is based on this observation. The higher-order terms do not produce significant or qualitatively different deviations compared to the lower-order terms. Furthermore, in the perturbative regime where \( j \ll 1 \), the dominant behavior of rotating AdS black holes is dictated by the leading-order terms, and higher-order terms remain subordinate. Attempts to associate new effects with these higher-order terms are constrained by their minimal impact on the overall results and their negligible deviation from the analytical predictions.

\section{Quintessence and string cloud impact on AdS black hole critical behavior}
\label{:4}
In this paragraph, our primary objective is to  explore  the intricate interplay between quintessence and a dense cloud of strings,  examining their combined influence on the critical behaviors exhibited by AdS black holes. 

To accomplish this, we create plots using Eq. (\ref{16}) to visualize the relationships between temperature and entropy, Gibbs free energy and temperature, and heat capacity and entropy. These plots are generated while varying the \(\omega_q\) parameter in the equation of state and the \( b \) parameter associated with the string cloud.

\begin{figure}[H]
	\includegraphics[width=1.05\linewidth, height=8.1cm]{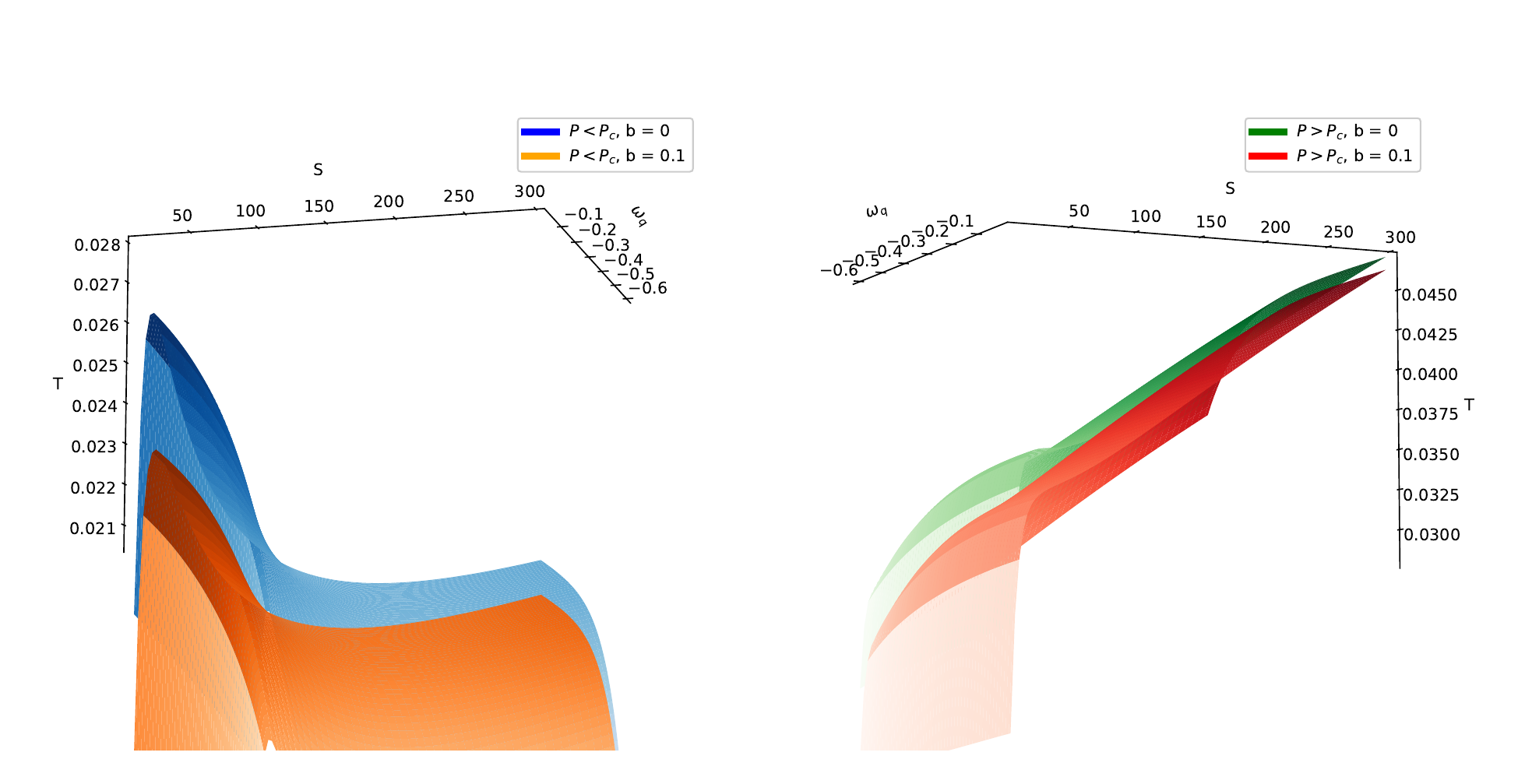}   
	\caption{ Temperature-entropy plot for charged rotating-AdS black holes with quintessence and cloud of strings: $ j = Q = 1 $, $ \alpha = 0.06 $. }
	\label{fig3}
\end{figure}
Figure \ref{fig3} illustrates the complex relationship between entropy and temperature for charged rotating AdS black holes that are immersed in an environment characterized by the presence of quintessence and a string cloud.

The temperature graph displays a fascinating oscillatory pattern characterized by two extreme points when the pressure $P$ is below the critical pressure $P_{C}$. However, as the pressure surpasses the critical threshold $P > P_{C}$, these extreme points vanish, and the oscillatory behavior dissipates, resembling the behavior of a van der Waals (vdW) system. Notably, as the parameter $\omega_{q}$ rises, the temperatures of the two extreme points shift towards higher values.

Furthermore, it is important to emphasize that, for varying values of $ \omega_q $, there is a significant reduction in both the phase transition point and the coexistence region of the two phases with an increasing parameter $ b $. This observation highlights the significant impact of the cloud of strings parameter on the thermodynamic characteristics of the system.

We create a graphical representation of the Gibbs free energy in relation to temperature, employing the provided expression for the Gibbs free energy. The specific formulas for $ G_{1} $, $ G_{2} $, $ G_{3} $, and $ G_{4} $ can be
found in the appendix.\
\begin{equation}
{G=\dfrac{S^{-\frac{1}{2}-\frac{3 \omega }{2}} \left(2 G_1^3-3 G_1^2 G_3+6 j^2 \pi ^2 S^{3 \omega } \left(2 G_2-3 G_4\right)\right)}{12
		\sqrt{\pi } G_1^2}}.\
\end{equation}
\begin{figure}[H]
	\centering
	\includegraphics[width=0.65\linewidth, height=10cm]{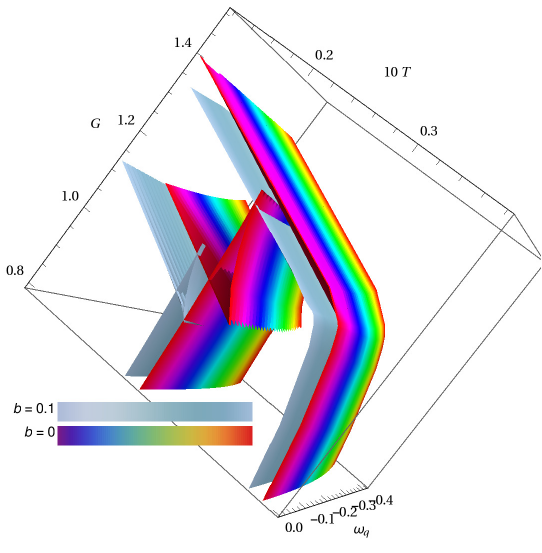}
	\caption{ Gibbs free energy of charged rotating-AdS black holes in the presence of quintessence and string clouds: $ j = Q = 1 $, $ \alpha = 0.06 $.\label{fig 31}}	
\end{figure}
The Gibbs diagram (Figure \ref{fig 31}) reveals a remarkable swallowtail pattern that stands out due to its unique shape. However, this distinctive swallowtail pattern only becomes apparent when the pressure $P$ is below the critical pressure $P_C$. Intriguingly, once the pressure reaches or exceeds the critical threshold ($P = P_C$ or $P > P_C$), this distinctive swallowtail pattern entirely vanishes from the diagram.

Clearly, as the parameter $b$ increases, the intersection point of the Gibbs points reflects a substantial downward trend. This trend correlates with a simultaneous rise in both Gibbs free energy $(G)$ and temperature $(T)$ throughout successive reductions in the parameter $b$. Importantly, this consistent and intriguing trend persists across a wide range of $\omega_q$ values, illustrating the robustness and universal applicability of this observed relationship in the realm of thermodynamics.

The formula representing the heat capacity of a black hole when it is in the presence of both a cloud of strings and quintessence can be expressed as such
 \

$ {C =T\dfrac{\partial S}{\partial T}=\left(2 S \lambda _1 \left(\lambda _1{}^2 \lambda _2+6 j^2 \pi ^2 S^{3 \omega _q} \lambda _3\right)\right)/\left(-24
	j^2 \pi ^2 S^{1+3 \omega _q} \lambda _3 \lambda _1'+\lambda _1{}^3 \left(-3 \left(1+\omega _q\right) \lambda _2+2 \right.\right.} $\

$ \hspace{2.5cm}{\left.\left.S \lambda _2'\right)+6 j^2 \pi ^2 S^{3 \omega _q} \lambda _1 \left(3 \left(-1+\omega _q\right) \lambda _3+2 S \lambda _3'\right)\right)}. $\

The specific formulas for $ \lambda_{1} $, $ \lambda_{2} $, and $ \lambda_{3} $ can be found in the appendix.

\begin{figure}[H]
	\centering
	\includegraphics[width=0.65\linewidth, height=9cm]{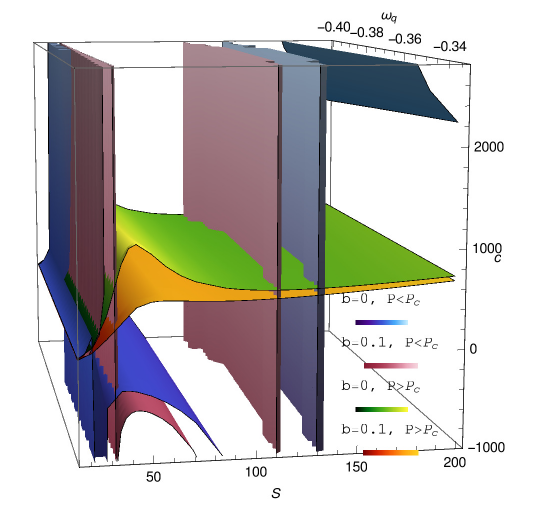}
	\caption{Heat capacity behavior of charged rotating-AdS black holes surrounded by quintessence and string clouds: $ j = Q = 1 $, $ \alpha = 0.06 $.} 
	\label{fig5}
\end{figure}

This heat capacity plot (Figure \ref{fig5}) reveals a distinct pattern influenced by the parameters associated with quintessence and a string cloud. When the pressure is below the critical value $ P < P_C $, the plot exhibits a negative slope on the ${T}-{S}$ curve, indicating the presence of an intermediate black hole branch with negative heat capacity, suggesting system instability. The divergence of the heat capacity signifies a phase transition in progress. Moreover, instability is also observed at $ S = S_c $ when the pressure equals the critical value $ P = P_C $, indicating a second-order phase transition. No phase transition occurs for pressures exceeding the critical value $ P > P_C $, leading to the restoration of system stability.
\section {Exploring fractional-order phase transitions in AdS black holes with quintessence and string clouds}
\label{:5}
Within the contents of this paper, we have expanded our investigation of the fractional order phase transition (FPT) occurring in a charged AdS black hole, particularly in the context of its interaction with quintessence and a cloud of strings.

For $ \omega_q = -\dfrac{2}{3} $, and using the critical point expressions obtained in section \ref{:3}, along with the newly introduced variables ($ t $, $ p $, $ v $), which are defined as follows
\[
P = (1 + p) p_c, \quad T = (1 + t) t_c, \quad V = (1 + v) v_c,
\]
the equation of state is formulated as follows
\[
P = \dfrac{4 Q^2 + V^2 (-1 + b + 2 \pi T V + V \alpha)}{2 \pi V^4},
\]
we find that the critical point is situated at ($ t = p = v = 0 $), and the equation of state simplifies to a quartic equation involving $ v $
\[
3 (-1 + b)^2 p (1 + v)^4 + (-1 + b)^2 v^3 (4 + 3 v) - 4 t (1 + v)^3 \left(2 - 4 b + 2 b^2 - 3 \sqrt{6 - 6 b} Q \alpha \right) = 0,
\]
once we've solved the earlier equation and simplified it, we adjust the solution in the following manner
\[
v = \dfrac{8 \cdot 6^{1/3} t^2 \left(-2 + 4 b - 2 b^2 + 3 \sqrt{6} \sqrt{1 - b} Q \alpha \right)^2}{27 (-1 + b)^4 p^{5/3}}.
\]
The rescaled Gibbs free energy takes the form
\[
g(t, p) = \dfrac{\sqrt{1 - b} \left(8 + 8v - 4v^3 - v^4 - p (1 + v)^4 \right)}{4 \sqrt{6} (1 + v)}.
\]
The equation previously mentioned can be effectively represented using the Taylor series with respect to $ t $, and it can be expressed in a more straightforward manner as follows
\[
g(t, p) = \Lambda(p) + \Gamma(p)t + D(p)t^2 + O[t]^3,
\]
with
\[
D(p) = -\dfrac{\left(2^{5/6} \sqrt{1 - b} \left(-2 + 4 b - 2 b^2 + 3 \sqrt{6} \sqrt{1 - b} Q \alpha \right)^2\right)}{9 \left(3^{1/6} (-1 + b)^4 p^{2/3}\right)}.
\]
As we approach the critical point, where both $t$ and $v$ are considerably smaller than 1, the equation of state can be simplified to the following expression
\[
p \approx \text{kt} + O\left[t^2, \text{tv}\right] = \dfrac{t_c t}{p_c v_c} = \left(\dfrac{8}{3} - 4 \sqrt{6} Q \alpha \right)t,
\]
in this context, we adopt the Caputo definition, which is particularly advantageous as it facilitates the convenient application of standard boundary and initial conditions~\cite{41}.
\[
D_t^{\beta} g(t) = \dfrac{1}{\Gamma (n - \beta)} \int_0^t (t - \tau)^{n - \beta - 1} \dfrac{\partial^n g(\tau)}{\partial \tau^n} \, \text{d}\tau, \quad n - 1 < \beta < n.
\]
Performing the calculation, taking into account $ \beta $ as the derivative order and considering the limit where both $ t $ and $ p $ tend toward zero, the result is expressed as follows
\[
D_t^{\beta} g(t, p)|_p = \dfrac{2 t^{2 - \beta}}{\Gamma (3 - \beta)} D(p), \quad t > 0; \ 1 < \beta < 2,
\]
\[
\lim_{t \to 0} D_t^{\beta} g(t, p)|_p = \begin{cases}
0 \quad \text{for} \quad \beta < \dfrac{4}{3}, \\
\dfrac{\left(2 - 4 b + 2 b^2 - 3 \sqrt{6 - 6 b} Q \alpha \right)^2}{\sqrt{6} (1 - b)^{7/2} \left(2 - 3 \sqrt{6} Q \alpha \right)^{2/3} \Gamma \left(\dfrac{2}{3}\right)} \quad \text{for} \quad \beta = \dfrac{4}{3}, \\
\infty \quad \text{for} \quad \beta > \dfrac{4}{3}.
\end{cases}
\]
For $ \omega_q = -\dfrac{1}{3} $, the equation of state is represented as
\[
P = \dfrac{4 Q^2 + V^2 (-1 + b + 2 \pi T V + \alpha)}{2 \pi V^4},
\]
using the recently introduced variables and solving the provided equation
\[
-8 t (1 + v)^3 + 3 p (1 + v)^4 + v^3 (4 + 3 v) = 0,
\]
we can find a solution to the equation as follows
\[
v = \dfrac{32 \cdot 2^{1/3} t^2}{9 \cdot 3^{2/3} p^{5/3}}.
\]
The rescaled Gibbs free energy takes the following form
\[
g(t, p) = \dfrac{\left(8 + 8v - 4v^3 - v^4 - p(1 + v)^4 \right) \sqrt{1 - b - \alpha}}{4 \sqrt{6} (1 + v)},
\]
using the Taylor series expansion, $ D(p) $ can be represented as
\[
D(p) = \dfrac{4 \cdot 2^{5/6} \sqrt{1 - b - \alpha}}{9 \cdot 3^{1/6} p^{2/3}}.
\]
As we approach the critical point, where both $ t $ and $ v $ become substantially less than 1, the equation of state can be simplified to the following expression
\[
p \approx \dfrac{t_c t}{p_c v_c} = \dfrac{8}{3} t,
\]
when we perform the computation, the final result is as follows

$$ \lim_{t \to 0} D_t^{\beta }g(t,p)|_p =\begin{cases}
0 \hspace{1cm}for \hspace{0.2cm} \beta < \dfrac {4}{3},  \\ 
{\text{}_+^- }{\dfrac{2^{5/6}\text{  }{\sqrt{1-b-\alpha }}}{\sqrt{3} \text{} \Gamma(\dfrac{2}{3})}}\hspace{1cm}for \hspace{0.2cm}  \beta = \dfrac {4}{3}, \\ {\text{}_+^-\infty }\hspace{1cm}for\hspace{0.2cm}\beta > \dfrac {4}{3}.
\end{cases}$$\

The research outcome provides strong evidence that, in the specific scenario under investigation, which involves a spherically symmetric AdS  black hole existing within the backdrop of a quintessence field and a surrounding cloud of strings, the nature of the fractional phase transition remains unchanged and consistently corresponds to a 4/3 order.

This fractional phase transition characterizes a critical transformation in the system, often revealing intriguing physical properties. The fact that it remains at the 4/3 order implies a particular resilience to external factors. In this case, the external factors include the presence of both the quintessence field and the cloud of strings that envelop the charged AdS black hole.
\section{Conclusion}
\label{:6}
This research examines the interactions between quintessence and a cloud of strings and their impact on critical phenomena in AdS black holes. The study focuses on how thermal fluctuations affect AdS black holes. The analysis shows that thermal fluctuations significantly influence smaller black holes, with the effect increasing as the corrected parameter $\beta$ grows across different values of $b$.
  Using  critical conditions, we identify the critical points for  various types of AdS black holes. These points are linked to the system’s parameters, $Q$ and $j$, as well as the $b$ parameter of the string cloud, the quintessential state parameter $\omega_q$, and the intensity parameter $\alpha$ representing the quintessential field surrounding the black hole.
A key finding of this study is the sensitivity of the system to the parameters $\omega_q$, $\alpha$, and $b$. As $\omega_q$ increases, the temperature increases, with the two extreme points shifting towards higher values. Moreover, changes in $\omega_q$ lead to a reduction in both the phase transition point and the coexistence region of the two phases. The intersection point of the Gibbs points also shifts towards higher temperature $T$ and higher Gibbs free energy $G$ with decreasing values of the parameter $b$.
Throughout the study, calculations consistently demonstrate that the fractional phase transition remains at an order of 4/3. After analyzing the critical points for  AdS black holes, the approach used in this study is validated, highlighting the combined effect of a string cloud and quintessence on the critical phenomena observed in AdS black holes.
\section*{{Acknowledgements}}
The authors would like to thank the anonymous referee for interesting comments and suggestions which motivated us to prepare a well-improved revised version.
\newpage
\appendix

\textbf{\Large{Appendix}}\

The expressions of the parameters used in the various sections of this work are presented as follows\

\noindent\({P_1=\left(-4 Q^2 v^{3 \omega _q}+(-1+b) v^{2+3 \omega _q}+2^{1+3 \omega _q} v \alpha \right){}^3 \left(v^{3 \omega _q} \left(4 Q^2+v^2
	(-1+b+2 \pi  T v)\right)-3\ 2^{1+3 \omega _q} v \alpha  \omega _q\right)},\)

\noindent\({P_2=\left(128 Q^6 v^{3 \omega _q}+16 Q^4 v \left(-v^{1+3 \omega _q}+4 \pi  T v^{2+3 \omega _q}-2^{2+3 \omega _q} \alpha \right)+v^5
	\left(\left(-8+11 b-3 b^2\right) v^{1+3 \omega _q}+4 \pi  T \right.\right.}\\
{\left.v^{2+3 \omega _q}-2^{1+3 \omega _q} (-4+b) \alpha +8^{1+\omega _q} \pi  T v \alpha \right)-4 Q^2 v^3 \left(\left(-1+2 b^2\right) v^{1+3 \omega _q}+4 (1+b)
	\pi  T v^{2+3 \omega _q}+2^{1+3 \omega _q}  \right.}\\
{\left.\left.(-1+2 b) \alpha+8^{1+\omega _q} \pi  T v \alpha \right)\right)},\)

\noindent\({P_3=\left(-32 Q^4 v^{3 \omega _q}+(-7+5 b) v^{4+3 \omega _q}+4 \pi  T v^{5+3 \omega _q}+2^{1+3 \omega _q} v^3 \alpha +4 Q^2 v \left((1+2
	b) v^{1+3 \omega _q}+2^{2+3 \omega _q} \alpha \right)\right) \omega _q},\)\

\noindent\( M_1c=\left(\sqrt{j} \left((-1+b+\alpha )^2\right)^{3/2} \left(-26418+26418 b+26418 \alpha -25610 \sqrt{(-1+b+\alpha )^2}+363 b \right.\right. \\
\left.\sqrt{(-1+b+\alpha )^2}+363 \alpha  \sqrt{(-1+b+\alpha )^2}\right)^2 \left(43879077 (-1+b+\alpha )^6 (26418+25610 \right. \\
\left.\sqrt{(-1+b+\alpha )^2}+3 b \left(4625-121 \sqrt{(-1+b+\alpha )^2}\right)+3 \alpha  \left(4625-121 \sqrt{(-1+b+\alpha )^2}\right)\right) \\
-\left((-1+b+\alpha )^2\right)^{3/2} \left(-26418-25610 \sqrt{(-1+b+\alpha )^2}+3 b \left(8806+121 \sqrt{(-1+b+\alpha )^2}\right)\right. \\
\left.\left.\left.+3 \alpha  \left(8806+121 \sqrt{(-1+b+\alpha )^2}\right)\right)^3\right)\right)/\left(1089 \sqrt{2} 37^{3/4} \left(\left((-1+b+\alpha
)^2\right)^{3/2}\right)^{1/4} \right. \\
\left(\left((-1+b+\alpha )^2\right)^{3/2} \left(-26418+26418 b+26418 \alpha -25610 \sqrt{(-1+b+\alpha )^2}+363 b \right.\right. \\
\left.\sqrt{(-1+b+\alpha )^2}+363 \alpha  \sqrt{(-1+b+\alpha )^2}\right)^2+1089 (-1+b+\alpha )^6 (13875-25610  \\
\sqrt{(-1+b+\alpha )^2}+3 b \left(-4625+121 \sqrt{(-1+b+\alpha )^2}\right)+3 \alpha  \left(121\sqrt{(-1+b+\alpha )^2}\right. 
{\left.-4625 )))^2\right)}, \)\

\noindent\({T_1=\left(-S^{\frac{3 \omega _q}{2}} \left(\pi  q^2+S (-1+b-8 P S)\right)+3 \pi ^{\frac{1}{2}+\frac{3 \omega _q}{2}} \sqrt{S} \alpha
	\omega _q\right)},\)

\noindent\({T_2=\left(3 \pi  Q^2 S^{\frac{3 \omega _q}{2}}+S^{1+\frac{3 \omega _q}{2}} (3-3 b+8 P S)-3 \pi ^{\frac{1}{2}+\frac{3 \omega _q}{2}}
	\sqrt{S} \alpha \right){}^2},\)

\noindent\({T_3=\left(-S^{\frac{3 \omega _q}{2}} \left(\pi  q^2 (3-8 P S)+S \left(b (-3+8 P S)+(3+8 P S)^2\right)\right)+48 P \pi ^{\frac{1}{2}+\frac{3
			\omega _q}{2}} S^{3/2} \alpha  \omega _q+9 \pi ^{\frac{1}{2}+\frac{3 \omega _q}{2}} \right.}\\
{\left. \sqrt{S}(3+8 P S) \alpha  \omega _q^2\right)},\)\

\noindent\(M_2c=\left(\left(3 \pi  Q^2-3 (-1+b+\alpha ) S_c+8 P_c S_c^2\right) \left(\left(3 \pi  Q^2-3 (-1+b+\alpha ) S_c+8 P_c S_c^2\right){}^5+12
j^2 \pi ^2 \right.\right.\\
S_c \left(3 \pi  Q^2-3 (-1+b+\alpha ) S_c+8 P_c S_c^2\right){}^2 \left(9 (b+\alpha )-128 P_c^2 S_c^2-24 P_c \left(2 \pi  Q^2-(-2+b+\alpha )
S_c\right)\right)\\
\left.\left.+9216 j^4 \pi ^4 P_c^2 S_c^2 \left(3 \pi  Q^2+S_c \left(3+8 P_c S_c\right)\right)\right)\right)/\left(6 \sqrt{\pi } \left(-96 j^2
\pi ^2 P_c S_c+\left(3 \pi  Q^2-3 (-1+b+\alpha )\right.\right.\right.\\
\left.\left.\left. S_c+8 P_c S_c^2\right){}^2\right){}^2 \sqrt{S_c \left(-12 j^2 \pi ^2 \left(3+8 P_c S_c\right)+\left(3 \pi  Q^2-3 (-1+b+\alpha
	) S_c+8 P_c S_c^2\right){}^2\right)}\right),\)\\	
\noindent\({\beta _1=\sqrt{Q^4-10 j^2 (-1+b+\alpha )} },\)\\
\noindent\({ \beta _2=\dfrac{\left(Q^2+\beta _1\right) \left(-Q^6+5 j^2 (-1+b+\alpha ) \left(3 Q^2-2 \beta _1\right)+Q^4 \beta _1\right)}{150 j^4 \sqrt{(-1+b+\alpha
			)^2}} },\)\\			
\noindent\({\beta _3=\dfrac{Q^2+\beta _1}{\sqrt{(-1+b+\alpha )^2}}},\)\\
$ { \sigma =\sqrt{-10 (-1+b) j^2 \pi ^3+\left(\pi ^{3/2} Q^2+2 \alpha \right)^2 } }, $\\
\noindent\(M_3c=\left(\left(3 \pi ^{3/2} Q^2-3 \alpha  S_c^{3/2}+\sqrt{\pi } S_c \left(3-3 b+8 P_c S_c\right)\right) \left(-36 j^2 \pi ^4 S_c
\left(3+8 P_c S_c\right) \left(288 j^2 \pi ^3 P_c S_c\right.\right.\right.\\ 
\left.-3\left(3 \pi ^{3/2} Q^2-3 \alpha  S_c^{3/2}+\sqrt{\pi } S_c \left(3-3 b+8 P_c S_c\right)\right){}^2\right)+9 \pi ^2 Q^2 \left(-96 j^2
\pi ^3 P_c S_c+\left(3 \pi ^{3/2} \right.\right.\\ 
\left.\left.Q^2-3 \alpha  S_c^{3/2}+\sqrt{\pi } S_c \left(3-3 b+8 P_c S_c\right)\right){}^2\right){}^2+S_c \left(-12 j^2 \pi ^3 \left(3+8 P_c
S_c\right)+\left(3 \pi ^{3/2} Q^2-3 \alpha  \right.\right.\\
\left.\left.S_c^{3/2}+\sqrt{\pi } S_c \left(3-3 b+8 P_c S_c\right)\right){}^2\right) \left(-288 j^2 \pi ^4 P_c S_c \left(3+8 P_c S_c\right)-9
\sqrt{\pi } \alpha  \epsilon  \left(3 \pi ^{3/2} Q^2-3 \alpha \right.\right.\\
\left. S_c^{3/2}+\sqrt{\pi } S_c \left(3-3 b+8 P_c S_c\right)\right)+3 \pi  \left(3-3 b+8 P_c S_c\right) \left(3 \pi ^{3/2} Q^2-3 \alpha  S_c^{3/2}+\sqrt{\pi
} S_c (3-3 b +\right. \\
\left.\left.\left.\left.\left.8 P_c S_c\right)\right){}^2\right)\right)\right)/\left(18 \pi ^{3/2} \epsilon  \left(-96 j^2 \pi ^3 P_c S_c+\left(3
\pi ^{3/2} Q^2-3 \alpha  S_c^{3/2}+\sqrt{\pi } S_c \left(3-3 b+8 P_c S_c\right)\right){}^2\right){}^2\right),\)\\
$ {\epsilon = \sqrt{S_c \left(-12 j^2 \pi ^3 \left(3+8 P_c S_c\right)+\left(3 \pi ^{3/2} Q^2-3 \alpha  S_c^{3/2}+\sqrt{\pi
		} S_c \left(3-3 b+8 P_c S_c\right)\right){}^2\right)}}, $\

\noindent\({c_1\text{$=$}\left(3 \pi  Q^2 S^{3 \omega _q /2}+S^{1+\frac{3 \omega _q }{2}} (3-3 b+8 P S)-3 \pi ^{\frac{1}{2}+\frac{3 \omega _q }{2}} \sqrt{S}
	\alpha \right)},\)\

\noindent\({c_2\text{$=$}3 \pi ^{\frac{1}{2}+\frac{3 \omega _q }{2}} \sqrt{S} \alpha  (9 \omega _q +8 P S (4+3 \omega _q ))},\)\

\noindent\({G_1=3 \pi  Q^2 S^{3 \omega /2}+S^{1+\frac{3 \omega }{2}} (3-3 b+8 P S)-3 \pi ^{\frac{1}{2}+\frac{3 \omega }{2}} \sqrt{S} \alpha
},\)\

\noindent\({G_2=3 \pi  Q^2 S^{3 \omega /2} (3+8 P S)+S^{1+\frac{3 \omega }{2}} \left(b (9-72 P S)+(3+8 P S)^2\right)-3 \pi ^{\frac{1}{2}+\frac{3
			\omega }{2}} \sqrt{S} \alpha  (9 \omega +8 P S (4+3 \omega ))},\)\

\noindent\({G_3=S^{3 \omega /2} \left(-\pi  Q^2+S (1-b+8 P S)\right)+3 \pi ^{\frac{1}{2}+\frac{3 \omega }{2}} \sqrt{S} \alpha  \omega },\)\

\noindent\({G_4=\pi  Q^2 S^{3 \omega /2} (-3+8 P S)-S^{1+\frac{3 \omega }{2}} \left(b (-3+8 P S)+(3+8 P S)^2\right)+3 \pi ^{\frac{1}{2}+\frac{3
			\omega }{2}} \sqrt{S} \alpha  \omega  (9 \omega +8 P S (2+3 \omega ))},\)\

\noindent\( {\lambda _1=\left(3 \pi  Q^2 S^{\frac{3 \omega _q}{2}}+S^{1+\frac{3 \omega _q}{2}} (3-3 b+8 P S)-3 \pi ^{\frac{1}{2}+\frac{3 \omega
			_q}{2}} \sqrt{S} \alpha \right)},\)\

\noindent\( {\lambda _2=\left(S^{\frac{3 \omega _q}{2}} \left(-\pi  Q^2+S (1-b+8 P S)\right)+3 \pi ^{\frac{1}{2}+\frac{3 \omega _q}{2}} \sqrt{S}
	\alpha  \omega _q\right)},\)\

\noindent\( \lambda _3= \left(-S^{\frac{3 \omega _q}{2}} \left(\pi  Q^2 (3-8 P S)+S \left(b (-3+8 P S)+(3+8 P S)^2\right)\right)+48 P \pi ^{\frac{1}{2}+\frac{3
		\omega _q}{2}} S^{3/2} \alpha  \omega _q\right. \\
\left.+9 \pi ^{\frac{1}{2}+\frac{3 \omega _q}{2}} \sqrt{S} (3+8 P S) \alpha  \omega _q^2\right) \).\

\end{document}